\documentclass[aps,prb,reprint,twocolumn,showpieces,superscriptaddress]{revtex4-2}
\usepackage{amsmath}
\usepackage{graphicx}
\usepackage{lmodern}
\usepackage{amsmath}
\usepackage{color}
\usepackage{amssymb}
\usepackage{bm}
\usepackage{braket}
\usepackage{mathtools}
\usepackage{afterpage}
\usepackage[normalem]{ulem}
\usepackage{soul}
\usepackage[caption=false,captionskip=-25pt, position=bottom]{subfig}

\usepackage[dvipsnames]{xcolor}
\usepackage[colorlinks=true]{hyperref}
\hypersetup{
  colorlinks=true,
  linkcolor=blue,
  citecolor=blue,
  urlcolor=blue
} 
\makeatletter
\def\maketitle{
\@author@finish
\title@column\titleblock@produce
\suppressfloats[t]}
\makeatother

\begin{document}
\title{
\texorpdfstring{
Enhanced Kohn-Luttinger superconductivity in geometric bands 
}{}}
\author{Ammar Jahin}
\affiliation{Theoretical Division, T-4, Los Alamos National Laboratory, Los Alamos, New Mexico 87545, USA}
\author{Shi-Zeng Lin}
\affiliation{Theoretical Division, T-4, Los Alamos National Laboratory, Los Alamos, New Mexico 87545, USA}
\affiliation{Center for Integrated Nanotechnologies (CINT), Los Alamos National Laboratory, Los Alamos, New Mexico 87545, USA}
\date{\today}

\begin{abstract}
We study the effect of the electron wavefunction on Kohn-Luttinger superconductivity. The role of the wavefunction is encoded in a complex form factor describing the topology and geometry of the bands. We show that the wavefunction significantly impacts the superconducting transition temperature and superconducting order parameter. We illustrate this using the lowest Landau level form factor and find exponential enhancement of $T_c$ for the resulting topological superconductor. {This enhancement of $T_c$ is due to the resonance between the Berry flux enclosed by the Fermi surface and Cooper pair angular momentum.} We find that the ideal band geometry, which favors a fractional Chern insulator in the flat band limit, has an optimal $T_c$. Finally, we apply this understanding to a model relevant to rhombohedral graphene multilayers and unravel the importance of the band geometry for achieving robust superconductivity.
\end{abstract}
\maketitle

\section{Introduction.} 

{Moir\'e superlattice and heterostructure of 2D materials emerged as an important platform to explore the interplay between strong correlations and topology, with superconductivity observed in twisted bilayer graphene~\cite{Cao_2018,Yankowitz_2019,Lu_2019,Chen_2019,Zhang_2024}, transition metal dichalcogenide (TMD)~\cite{Xia_Han_Watanabe_Taniguchi_Shan_Mak_2024,Guo_Pack_Swann2024}, and graphene multilayers~\cite{Zhou_2021,Zhou_2022,Zhang_2023,Li_2024}. }
It is common for superconductivity to arise from the pairing between time-reversal partners, i.e., electrons at opposite valleys. 
Recently, superconductivity emerging from a spin and valley-polarized normal state has also been observed in tetralayer graphene~\cite{Long_2024}.
While the mechanism for superconductivity in these systems remains under debate~\cite{PhysRevLett.122.026801,PhysRevB.98.195101,Chou_2021,Li_2022,Ghazaryan_2021,You_2022,Chatterjee_2022,Sarma_2024,Fu_2024,Hsu_2017,Wang_Gao_Yang_2024,Yang_Zhang_2024,Jimeno_2023,Li_2023,Dong_2023,Dong_2024,Cea_2022,Chou_2022,Dong_2023_b,Guinea_2021}, one appealing proposal is the Kohn-Luttinger (KL) mechanism~\cite{Kohn_1965} where the pairing potential arises merely from the repulsive Coulomb interaction. 
{Earlier work has indicated that the electronic wavefunction on the Fermi surface can affect KL mechanism~\cite{Ammar_2023}}. 
Motivated by these exciting developments, we study the role of the electron wavefunction on KL pairing $T_c$.

 For theoretical description, it is convenient to project the Hamiltonian into a reduced Hilbert space associated with the bands near the Fermi surface. 
After projection, the wavefunction enters the Hamiltonian through the form factor $\Lambda_{\tau}(\bm k, \bm q) = \langle u_{\tau}(\bm k)|u_{\tau}(\bm k+\bm q) \rangle$ with $|u_{\tau}(\bm k)\rangle$ being the periodic part of the Bloch wavefunction {with $\tau$ describing the valley degree of freedom}. 
If all we care about is, for example, the superfluid stiffness, defined as a small twist of the superconducting order parameter phase then it is sufficient to consider a small $\bm q$ limit, where the quantum geometric tensor describes the norm of the form factor. ~\cite{Peotta_2015,Julku_2016,Liang_2017,Hu_2019,Chen_2024,Tian_2023,Julku_2020,Xie_2020,He_2021,Herzog_2022,T_rm_2022,Torma_2018,Huhtinen_2022,Chen_2023,Wang_2025,Chen_2024,Hu_2024,Ziting_2024,PhysRevLett.131.016002}
However, in the KL mechanism, excitation processes with large momentum transfer $\bm q$ are also important for screening the Coulomb interaction. 
This requires keeping the full form factor $\Lambda(\bm k, \bm q)$, not only a small $\bm q$ expansion. 
In general, the KL mechanism generates extremely low $T_c$. 
One way to enhance $T_c$ is to tune the system close to a van Hove singularity~\cite{Nandkishore_2014}. 
{The central question is whether band geometry can offer an alternative route.}  

In this work, we consider the case of the pairing of electrons in a spin and valley-polarized metal without time-reversal symmetry. 
We partition the form factor corrections to the pairing vertex into its norm $|\mathcal W|$ (defined later), which is gauge invariant, and a gauge-dependent part $\exp(iF)$. 
The norm $|\mathcal W| \leq 1$ is related to the quantum distance in the Hilbert space and controls the screening of the bare Coulomb interaction, as well as act as an overall reduction factor to pairing vertex. 
The former effect {increases} $T_c$, while the latter {decreases it}. 
Interestingly, we show that the phase of the form factor splits the degeneracy in the pairing channels with opposite angular momentum, and tends to enhance $T_c$. 
We introduce a toy model with the lowest Landau level (LLL) form factor to illustrate the enhancement of $T_c$ by orders of magnitude. 
We also identify resonance in $T_c$ associated with the total Berry flux enclosed in the Fermi surface. 
As a comparison, we compute $T_c$ in a valley unpolarized system with time-reversal symmetry, where only $|\mathcal W|$ enters into the gap equation. 
The critical temperature, in this case, is lower, and in some regions, can be lower by orders of magnitude compared to the case without time-reversal symmetry. 
Finally, we apply our results to a two-band model describing rhombohedral multilayer graphene.

\begin{figure}
    \centering
    \captionsetup[subfigure]{oneside,margin={-8cm,0cm}, captionskip=-40pt}
    \subfloat[]{\includegraphics[scale=0.62, trim = 0 100 0 0, clip]{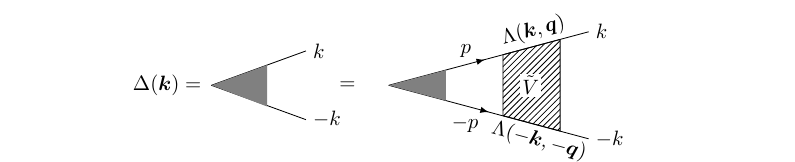}}

    \captionsetup[subfigure]{oneside,margin={-8cm,0cm}, captionskip=-70pt}
    \subfloat[]{\includegraphics[scale=0.62, trim = 0 0 0 90, clip]{figures/valley_polarization.pdf}}
    \caption{(a) Linearized gap equation including the effects of band geometry as captured by the form factor $\Lambda(\bm k, \bm q)$. 
    (b) The difference between intervalley pairing and intravalley pairing. The form factor adds a correction to the pairing vertex that is real because of time-reversal symmetry for intervalley pairing. However in a valley and spin-polarized metals, time-reversal symmetry is broken in the normal state, and the form factors correction is generally complex.}
    \label{fig:illustration}
\end{figure}

\section{Gap equation.}

The band projected Hamiltonian is 
\begin{align}
    H = \sum_{\bm k} (\varepsilon(\bm k) - \mu) a^\dagger_{\tau, \bm k} a_{\tau, \bm k} + \frac{1}{2L^2}\sum_{\bm q} V(\bm q) \tilde \rho(\bm q) \tilde \rho(-\bm q),
\end{align}
where $a_{\tau \bm k}$ annihilate an electron in the band of interest, $\tau = \pm$ is the valley degree of freedom, and $\tilde \rho(\bm q) = \sum_{\tau, \bm k} \Lambda_\tau(\bm k, \bm q) a^\dagger_{\tau, \bm k} a_{\tau, \bm k + \bm q}$ is the projected density operator and $L^2$ is the area of the system. {We neglect scattering between the two valleys.}
We will consider systems with the valley degree of freedom being either fully polarized or unpolarized. 
We assume the system is either fully spin-polarized on each valley or spin is slaved to the valley as in TMD. 
We use a gate-screened Coulomb interaction $V(\bm q) = 2\pi e^2 \tanh(q d) / \epsilon q$, where $d$ is the distance between the gates and the 2D system, and $\epsilon$ is dielectric permittivity.
We use $d = 36.9 \ \text{nm}$ and $\epsilon = 4 \epsilon_0$, where $\epsilon_0$ is the vacuum permittivity.  

To describe the screening of the bare repulsive interactions between the electrons by the particle-hole fluctuations at the Fermi surface, we use a random phase approximation (RPA), 
\begin{align}
    \tilde V(\bm q) = \frac{V(\bm q)}{1 + \Pi(\bm q) V(\bm q)},
\end{align}
where $\Pi(\bm q)$ is the static polarization {bubble} described by the Lindhard function with the vertices modified by the form factor of the interactions~\cite{SM},
\begin{align}\label{eq:bubble_w_ff}
    \Pi(\bm q) = - N \int d\bm k  \ |\Lambda(\bm k, \bm q)|^2 \frac{f(\varepsilon(\bm k + \bm q)) - f(\varepsilon(\bm k))}{\varepsilon(\bm k + \bm q) - \varepsilon(\bm k)}, 
\end{align}
where $N$ is the number of occupied valleys, i.e. $N=1$ for valley polarized systems and $N=2$ for valley unpolarized systems, and $f(\varepsilon)$ is the Fermi-Dirac distribution function. 

{The gap function is defined as $\Delta_{\tau,\tau'}(\bm k) = - \sum_{\bm k} \mathcal{W}_{\tau, \tau'} (\bm k, \bm k') \tilde V(\bm k - \bm k') \langle c_{\tau, -\bm k}  c_{\tau', \bm k} \rangle$, where $\mathcal W(\bm k, \bm k') = \Lambda_{\tau}(-\bm k, -\bm k' + \bm k) \Lambda_{\tau'}(\bm k, \bm k' - \bm k) $}. 
{Intervalley (intravalley) pairing is defined as the case when $\tau \neq \tau'$ ($\tau = \tau'$)}
{We note that in the intervalley pairing case with time-reversal symmetry, $\Lambda_+(\bm k, \bm k' - \bm k) = \Lambda_-^*(-\bm k, -\bm k' + \bm k)$, and $\mathcal W_{\tau, \tau'}(\bm k, \bm k')$ is real and positive.
However, for intravalley pairing, this does not necessarily hold, which we highlight pictorially in Fig.~\ref{fig:illustration} (b). 
As we will discuss, this can have big impact on superconductivity. 
In the following, we focus on the more interesting intravalley case, and drop the valley label for brevity. We come back to comment on intervalley pairing later. 
}
The linearized gap equation is sketched in Fig.~\ref{fig:illustration} (a) takes the following form,
\vspace{-6pt}
\begin{align}\label{eq:linearized_gap_eq}
    &\Delta(\bm k) = \\  &- \log\left(\frac{W}{T_c} \right)  \int_{\text{FS}} \frac{dk'_{\parallel}}{(2\pi)^2 v(\bm k')} \mathcal W(\bm k, \bm k') \tilde V(\bm k - \bm k') \Delta(\bm k') \nonumber 
\end{align}
where $W$ is the upper energy cut off of the integral, which is of the order of the Fermi energy, and $v(\bm k') = |\partial \varepsilon(\bm k') / \partial \bm k'|$. 

We focus on systems with a rotational symmetry along the axis perpendicular to the 2D plane such that both the dispersion and the form factor are invariant under continuous rotations in the plane. We then can write the gap equation in terms of the angles $\theta$ ($\theta'$) between $\bm k$ ($\bm k'$) and the $x$-axis. 
We rewrite $\mathcal W(\theta - \theta') = |\mathcal W(\theta - \theta')| e^{iF(\theta - \theta')}$, and define, 
\begin{align}
    \mathcal V (\theta) = |\mathcal W (\theta)| \tilde V(\theta)
\end{align} 
such that the gap equation takes the form
\begin{align}\label{eq:gap_eq_theta}
    \Delta(\theta) = - \log\left(\frac{W}{T_c} \right)   \frac{k_f}{v2\pi} \int  \frac{d\theta'}{2\pi} \  e^{iF(\theta - \theta')}  \mathcal V(\theta - \theta') \Delta(\theta')
\end{align}
where we have used $d k'_{\parallel} = d\theta' k_f $. 
We note that $ |\mathcal W(\theta - \theta')|$ is controlled by the quantum distance between states in the Fermi surface, which is captured by the Fubini-study metric for $|\theta - \theta'|\ll 1$, whereas the phase  $e^{iF(\theta - \theta')}$ is related to the Berry phase gained as we parallel transport the Bloch wavefunction along the Fermi surface. 

Decomposing the gap equation into its angular momentum components, we have 
\begin{align}
    \Delta_l = \log\left(\frac{W}{T_{lc}} \right) \lambda_l \  \Delta_l
\end{align}
where $T_{lc}$ is the critical temperature for the $l$ angular momentum channel and we define 
\begin{align}
    &\Delta_l = \int \frac{d\theta}{2\pi} e^{-il\theta} \Delta(\theta) \\  
    &\lambda_l = -\frac{k_f}{v2\pi} \int \frac{d\theta}{2\pi} e^{i(F(\theta) - l \theta)} \mathcal V(\theta). \label{eq:lambda_l_def}
\end{align}
{The true critical temperature is $T_c = \text{max}(T_{lc})$.}
In cases with time-reversal symmetry, $F(\theta) = 0$ and we have $\lambda_l = \lambda_{-l}$. 
The case with $F(\theta) \neq 0$ breaks the degeneracy, enhancing one channel while reducing the other, leading to an overall enhancement to $T_c$.
In fact, in what follows, we describe a situation where the full behavior of the system is governed by the phase factor $F(\theta)$. 
In general, the result of the convolution leading to maximal $\lambda_l$ is highly nontrivial, since both $e^{iF(\theta)}$ and the $\mathcal{V}(\theta)$ will have non-zero components on a wide range of angular momenta.

\begin{figure}
    \centering
    \begin{minipage}{0.217\textwidth}
    \subfloat[$\qquad \qquad \quad $]{%
    \includegraphics[scale=0.6, trim = 0 217 5 0, clip]{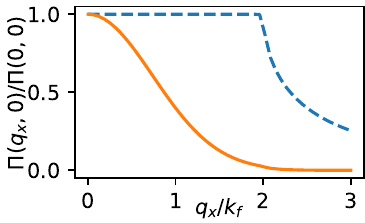}}
    \vspace{-10pt}

    \subfloat[$\qquad \qquad \quad $]{%
    \includegraphics[scale=0.6, trim = 0 113 5 108, clip]{figures/renorm_int.pdf}}
    \vspace{-10pt}
    
    \subfloat[$\qquad \qquad \quad $]{%
    \includegraphics[scale=0.6, trim = 0 3 5 214, clip]{figures/renorm_int.pdf}}
    \vspace{-8pt}
    \end{minipage}%
    \hfill
    \begin{minipage}{0.265\textwidth}
        \centering 
        \captionsetup[subfigure]{oneside,margin={4.3cm,0cm}, captionskip=-10pt}
        \subfloat[]{\includegraphics[scale=0.6, trim = 3 190 3 0, clip]{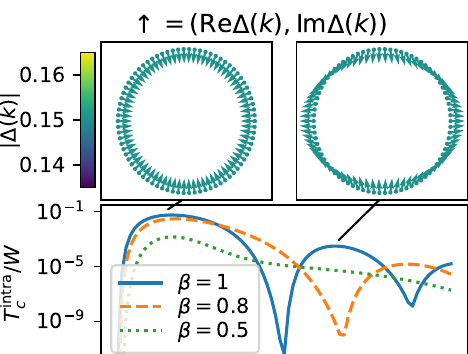}} 
        \vspace{-12pt}
        \captionsetup[subfigure]{oneside,margin={4.3cm,0cm}, captionskip=-18pt}
        \subfloat[]{\includegraphics[scale=0.6, trim = 3 100 3 170, clip]{figures/toy_model_tc_inter_intra_bessel.pdf}} 
        \vspace{-14pt}
        \captionsetup[subfigure]{oneside,margin={4.3cm,0cm}, captionskip=-31pt}
        \subfloat[]{\includegraphics[scale=0.6, trim = 3 0 3 255, clip]{figures/toy_model_tc_inter_intra_bessel.pdf}} 
    \end{minipage}
    \caption{Effect of band geometry on the bubble (a) and renormalized interactions (b). Panel (c) shows the Fourier transform of the renormalized interactions for different angular momentum channels for the cases with trivial band geometry $|\mathcal W|=1$ (squares), and non-trivial band geometry (circles). Non-trivial band geometry gives rise to an attractive channel that is otherwise absent. In (d) we plot $T_c$ for the case of intravalley pairing in units of $W$, the energy cutoff, for various values of $\beta$. $\beta = 1$ corresponds to the ideal band geometry limit, which provides the highest $T_c$. The corresponding superconducting order parameters at the Fermi surface are displayed with arrows representing the phase of the order parameter. Panel (e) compares the critical temperatures of intervalley pairing with intravalley pairing, and the intravalley pairing is significantly favored.
    We can analytically show in the case of large electron mass that the coupling constant for any odd pairing channel is given by the Bessel function, and panel (f) shows that it agrees well with the numerical results of pannel (d). }
    \label{fig:rpa_results}
    \vspace{-10pt}
\end{figure}

The form factor appears in two places in the gap equation. 
It changes the polarization bubble at non-zero momentum transfer, which modifies the renormalization of the interactions. 
This effect leads to a situation where the energy cost for forward scattering is suppressed compared to the energy cost for scattering with a large momentum transfer, see Fig.~\ref{fig:rpa_results} (b).
This means that pairs prefer to ``not scatter" rather than scatter off each other, which enhances their chances of forming pairs. 
However, the effects of band geometry also show up as an overall factor $\mathcal W(\theta)$ in the pairing vertex as seen in Eq.~\eqref{eq:linearized_gap_eq}. 
Since $|\mathcal W(\theta)| \le 1$, this leads to suppression of the critical temperature. 
This highlights a competition between these two effects, and which one wins depends on the details of the system. 

\section{Lowest Landau level form factor}

The most striking example of the effects of the band geometry is where we have parabolic dispersion $\varepsilon(\bm k) = \bm k^2/(2m)$ in 2D, where the conventional KL mechanism cannot lead to a superconducting instability~\cite{Kagan_2015}.  
The reason is that the polarization bubble is constant $\Pi(\bm q) = N m/(2\pi)$ for all momentum transfers $q \le 2k_f$, and the singularity in the bubble is single-sided as shown by the dashed line in~\ref{fig:rpa_results} (a).
It has been shown that this behavior can change if we include trigonal warping effects in the dispersion~\cite{Baranov_1992}, or corrections to screening up to third order in perturbation theory~\cite{Chubukov_1993}. 
Here we show that the LLL form factor, which has the form
\begin{align}\label{eq:lll_formfactor}
    \Lambda_{\text{LLL}}(\bm k, \bm q) = \exp\left[-\frac{B}{4} (q^2 + 2\beta i  \bm q \times \bm k) \right],  \qquad \beta \le 1,
\end{align}
generates high $T_c$ in the system. Here we introduce $\beta \leq 1$ to allow the system to deviate from the LLL limit ($\beta=1$). {In particular, $\beta$ controls the strength of the Berry curvature.}
The LLL form factor has the property that $|\Lambda_{\text{LLL}}(\bm k, \bm q)|$ does not depend on $\bm k$, which allows for analytical progress. 
We can immediately show from Eq.~\eqref{eq:bubble_w_ff} that the bubble is not a constant to $2k_f$, but rather $\Pi(\bm q) \propto e^{-Bq^2/2}$ as shown in Fig.~\ref{fig:rpa_results} (a). 
This has a huge effect on the renormalized interactions between electrons at the Fermi surface. 
In Fig.~\ref{fig:rpa_results} (b) we plot the renormalized interaction with and without the form factor in solid and dashed lines respectively.
Due to the form factor, scattering with high momentum transfers costs more energy than forward scattering, creating attractive channels as shown in Fig.~\ref{fig:rpa_results} (c).
Since we are using a parabolic dispersion, the band geometry guarantees to be favorable for superconductivity.

In Fig.~\ref{fig:rpa_results} (d) we show $T_c$ in units of the energy cutoff $W$ in the case of intravalley pairing. 
We see clear resonances to $T_c$ as we increase $B$. 
In the insets, we plot the order parameter around the Fermi surface, for the first two peaks of the ideal limit $\beta = 1$ plot, showing that the leading instabilities are for angular momentum channels $l = 1$ and $l=-1$ respectively at these points. {These are chiral $p_x\pm ip_y$ superconductors.}
We also show $T_c$ as we vary $\beta$ and deviate from the ideal geometry condition at $\beta=1$. 
Having ideal band geometry is also ideal for superconductivity, yielding the highest $T_c$. 

Next, we ask how the effect of band projection differs between intravalley and intervalley pairing. 
There are two differences between these cases: first, $\mathcal W(\theta)$ is complex for intravalley pairing, and real for intervalley pairing, and second, in Eq.~\eqref{eq:bubble_w_ff}, $N = 1$ for intravalley, and $N=2$ for intervalley pairing. 
While both intervalley and intravalley pairings have the same qualitative change in the bubble behavior, the intravalley pairing gets a much greater enhancement to $T_c$ as shown in Fig.~\ref{fig:rpa_results} (e). 
The reason is that the complex phase factor $e^{iF(\theta)}$ allows for a great enhancement to $T_c$ in the case of intravalley pairing. 
{We note that the ratio $T_c^{\text{intra}} /T_c^{\text{inter}} $ diverges as $Bk_f^2$ goes to zero. This divergence is to be understood that the intervalley pairing problem has no superconducting instability for small values of $Bk_f^2$.}
{That intravalley is favored over intervalley is consistent with the fact that the ideal geometry, $\beta = 1$ gives the highest $T_c$. 
Tuning $\beta$ between $1$ and $0$ can be thought of as smoothly interpolating between the two limiting cases of intravally pairing and intervalley pairing, since in the case of $\beta = 0$, $\mathcal W (\theta)$ is again real and takes the same form as for intervalley pairing.}

\section{Resonance between Berry curvature and Cooper pairs angular momentum}
The results with LLL from factor and parabolic dispersion can be analytically understood in the limit where the mass $m$ is large such that $e^{-Bq^2/2}V(\bm q)\  m /(2\pi) \gg 1$. Then we can approximate $\mathcal V(\bm q) = 2\pi/ m $. The coupling constant for every channel {in the case of intravalley pairing} is simply given by
\begin{align} \label{eq:lambda_resonance}
    \lambda_l = -\int \frac{d\theta}{2\pi} e^{-i ( \beta Bk_f^2 \sin(\theta) +  l\theta)}. 
\end{align}
The above expression has an analytical form in terms of Bessel functions, which for odd channels relevant for intravalley pairing takes the form, 
\begin{align}\label{eq:lambda_bessel}
    \lambda_l = J_{l}(\beta Bk_f^2), \qquad l \text{ is odd}. 
\end{align}
The highest $T_c$ is given by $Bk_f^2$ corresponding to the maximal of Bessel functions, which is not the simple quantization condition for the total Berry flux enclosed in the Fermi surface, $B \pi k_f^2 = 2\pi n$, as one would naively expect. We can also make analytical progress in two other limiting cases as described in~\cite{SM}. 

{Eq.~\eqref{eq:lambda_resonance} can be understood as a resonance between mainly the Berry flux enclosed by Fermi surface in the normal state, and the Cooper pair wavefunction as follows. 
The superconducting wavefunction is given by
\begin{align}
    | \Psi_s \rangle = \prod'_{\bm k} \left(u_{\bm k} + v_{\bm k} a^\dagger_{-\bm k} a^\dagger_{\bm k} \right) | 0 \rangle
\end{align}}
{where the prime superscript on the product indicates that every momentum pair $\{\bm k,-\bm k \}$ give only one factor in the product, $| 0 \rangle$ is the vacuum, and we drop the valley index from the creation operators as we only focus on intravalley pairing. 
Without loss of generality we take $u_{\bm k}$ to be real, but $v_{\bm k}$ can be complex. 
The expectation value of the energy for the superconducting state is given by 
\begin{align}
    \langle \Psi_s | H | \Psi_s \rangle = \langle H_0 \rangle + \langle H_{\text{int}} \rangle =  \sum_k \left( \varepsilon(\bm k) -\mu \right) |v_{\bm k}|^2 \nonumber \\ 
    + \frac{1}{2L^2}\sum_{\bm k, \bm k'} u_{\bm k'} v^*_{\bm k'} \  \mathcal W (\bm k', \bm k) \tilde V(\bm k' - \bm k) \  u^*_{\bm k} v_{\bm k}
\end{align}
where $\langle H_0 \rangle$ represents the expectation value of the kinetic term, and $\langle H_{\text{int}} \rangle$ is the expectation value of the interaction potential.
}

{The phase of $v_k$ varies as we go around the Fermi surface. We assume an SO($2$) rotational symmetry, so that different angular momentum channels do not mix. The solution minimizing the superconducting state energy takes the form 
\begin{align}
    | \Psi^l_s \rangle = \prod'_{\bm k} \left(u^l_{\bm k} + |v^l_{\bm k}| e^{i l \theta(\bm k)} a^\dagger_{-\bm k} a^\dagger_{\bm k} \right) | 0 \rangle
\end{align}
where $\theta(\bm k)$ is the angle $\bm k$ make with the $x$-axis. 
As before, $u^*_{\bm k} v_{\bm k}$ is nonzero only within a small energy window $W$ around the Fermi surface. Inside this energy window, we can take both $\mathcal W (\bm k', \bm k)$ and $\tilde V(\bm k'-\bm k)$ to only depend on the angle between $\bm k$ and $\bm k'$. 
Evaluating the interaction term for each angular momentum we obtain 
\begin{align}
    \frac{\langle H_{\text{int}} \rangle}{L^2} = \frac{1}{2} \left(\int_{k_f - k_W}^{k_f + k_W} d k  \ k |u^l_k| |v^l_k| \right)^2  \nonumber \\ 
    \times \int \frac{d\theta}{2\pi} \frac{d\theta'}{2\pi} e^{il(\theta - \theta')} \mathcal W(\theta - \theta') \tilde V(\theta - \theta') \nonumber \\ 
    = \frac{1}{2} \left(\int_{k_f - k_W}^{k_f + k_W} d k  \ k \  |u^l_k| |v^l_k| \right)^2  
    \int \frac{d\theta}{2\pi}  e^{il\theta } \mathcal W(\theta ) \tilde V(\theta ). 
\end{align}
where $k_W$ is the momentum cut-off coresponding to the energy cut-pff $W$. 
Minimizing the energy with respect to the magnitudes $|u^l_{\bm k}|$ and $|v^l_{\bm k}|$ is the same for all angular momentum channels, i.e. these magnitudes do not depend on the angular momentum. 
However, the reduction (or increase) in the interaction energy is dictated by the integral
\begin{align}
    \alpha_{l} = - \int \frac{d\theta}{2\pi}  e^{il\theta } \mathcal W(\theta ) \tilde V(\theta ) 
\end{align}
which is related to the coupling $\lambda_l$ defined in Eq.~\eqref{eq:lambda_l_def}. 
For the superconducting state to have a lower energy than the normal state, we need $\alpha_l > 0$. Furthermore, the angular momentum channel with the largest $\alpha_l$ has the lowest energy.
}

{The condition of $\text{max}(\alpha_l)$ can be understood as a resonance between the Cooper pair wavefunction $\psi(\bm k) = \langle c_{\bm k} c_{-\bm k} \rangle = u^*_{\bm k} v_{\bm k}$ and the factor ${\mathcal W(\theta )} \tilde V(\theta)$ which is the product of the form factors dictated by the quantum geometry of the system, and the screened Coulomb interactions (also modified by band geometry). 
From this point of view, it is natural that the Berry flux enclosed by the Fermi surface has the biggest effect on such resonance, being the only effect that can introduce chirality to the system, favoring a certain angular momentum channel. 
As the Berry flux in the system is increased, the winding in the phase of form factor increases, leading to higher angular momenta being in resonance.
In a system with a parabolic dispersion with the form factors taken to mimic those of the LLL~\eqref{eq:lll_formfactor}, we have that, in the large mass limit, the resonance condition is nothing but Eq.~\eqref{eq:lambda_bessel}. 
}

{Repeating the same exercise for intervalley pairing one reach the conclusion that superconductivity does not develop for all channels, due to the fact that the phase of the from factors cancel out, leaving the kernel to be a constant in this case. 
This leads to a sinuation where $\lambda_{l} = 0$ for all $l \neq 0$ while $\lambda_{l=0} < 0$, as the effects of screening is completely canceled by the amplitude of form factors modifying the pairing vertex. 
This support our numerical results that quantum geometry favors intravalley pairing.}

\begin{figure}
    \centering
    \begin{minipage}{0.237\textwidth}
    \captionsetup[subfigure]{oneside,margin={-1.4cm,0cm}, captionskip=-30pt}
    \subfloat[]{\includegraphics[scale=0.675, trim = 0 0 182 0, clip]{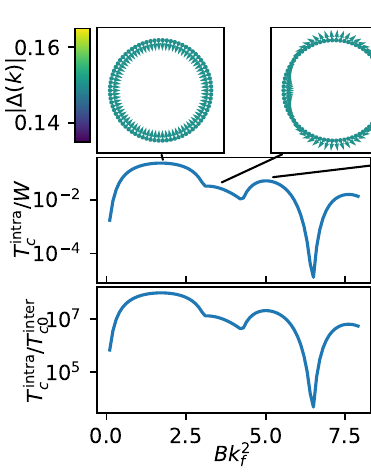}}
    \end{minipage}
    \hspace{-6.8pt}
    \begin{minipage}{0.239\textwidth}
    \captionsetup[subfigure]{oneside,margin={3.5cm,0cm}, captionskip=-30pt}
    \subfloat[]{\includegraphics[scale=0.675, trim = 177.7 0 2 0, clip]{figures/quartic_tcs_w_vecs.pdf}}
    \end{minipage}
    \vspace{-10pt}
    \caption{Superconducting critical temperature using quartic dispersion. The top panels of (a) and (b) show the critical temperature of intravalley pairing $T^{\text{intra}}_{c}/W$ and intervalley $T^{\text{inter}}_{c}/W$ respectively as a function of $B$. Intravalley pairing has a higher critical temperature for all values of $B$. The bottom panels show the same quantities but as a ratio of $T^{\text{inter}}_{c0}$ which is the critical temperature of intervalley pairing with trivial band geometry $|\mathcal W|=1$. The case of intravalley pairing with {the LLL band geometry} always leads to an enhanced $T_c$. However, for the intervalley pairing, the effect of band geometry can suppress $T_c$. Top row displays the corresponding superconducting order parameter at Fermi surface.}
    \label{fig:quartic_LLL}
    \vspace{-10pt}
\end{figure}

We caution that the enhancement of $T_c$ by the band geometry is not universal, which can be illustrated for a quartic band dispersion $\epsilon(\bm k)=\gamma \bm k^4$, where $\gamma$ is an energy scale. In Fig.~\ref{fig:quartic_LLL} (a) we show the critical temperature of intravalley pairing, as a ratio of the energy cutoff $T^{\text{intra}}_c / W$ (top) and as a ratio of the critical temperature of intervalley pairing with a trivial form factor $B=0$, $T^{\text{intra}}_c / T^{\text{inter}}_{c0}$ (bottom). 
We see a significant enhancement in $T^{\text{intra}}_c$ compared to $T^{\text{inter}}_{c0}$. 
This enhancement is mainly attributed to the phase factor $e^{iF}$. 

For intervalley pairing, we have $F = 0$, and similar to the quadratic dispersion case, this reduces the critical temperature compared to the intravalley pairing.
In Fig.~\ref{fig:quartic_LLL} (b) (top) we plot the intervalley pairing critical temperature as a ratio of the energy cutoff $T^{\text{inter}}_c / W$, which shows clearly $T^{\text{inter}}_c \ll T^{\text{intra}}_c$. 
Meanwhile, comparing $T^{\text{inter}}_c$ to intervalley pairing with a trivial form factor, we do not get an enhancement of $T_c$ for all values of $B$ as shown in Fig.~\ref{fig:quartic_LLL} (b) (bottom). 
This is due the $\mathcal W$ factor in the gap equation~\eqref{eq:linearized_gap_eq}. Since  $|\mathcal W| \leq 1$, it acts to reduce $T_c$ compared to the cases with no band geometry. 
We emphasize that in other cases where we do find enhancement, it is because of the effect of band geometry in the bubble showing up in Eq.~\eqref{eq:bubble_w_ff}, which tends to increase $T_c$. 
This situation shows that the result of this competition is not universal, and in some scenarios, the band geometry can be bad for superconductivity.

\section{Two-band model}
A two-band model for the low-energy physics of rhombohedral $n$-layer graphene systems can be written as~\cite{PhysRevB.82.035409,PhysRevB.109.205122,Slizovskiy_2019}
\begin{align}\label{eq:two_band_Hamiltonian}
    H(\bm k) = \begin{bmatrix}
        D && \gamma (k_x + i k_y)^n \\ 
        \gamma (k_x - i k_y)^n && -D 
    \end{bmatrix} - \mu,
\end{align}
where $D$ is a perpendicular displacement field, and $\gamma$ is an energy scale. 
The basis of this Hamiltonian is the $A$ sublattice of the top layer and the $B$ sublattice of the bottom layer. 
This Hamiltonian describes a system of electrons with a dispersion $\varepsilon(\bm k) = \sqrt{\gamma^2 k^{2n} + D^2}$ and the form factor between $\bm k$ and $\bm k'$ on the Fermi surface as, 
\begin{align}
    \Lambda(\bm k, \bm k' - \bm k) = \cos^2(\theta/2) + \sin^2(\theta/2) e^{in\phi}.
\end{align}
where, $\phi$ is the relative angle between $\bm k$ and $\bm k'$ and $ \cos(\theta) = D/\mu $. 

\begin{figure}
    \centering
    \captionsetup[subfigure]{oneside,margin={3.5cm,0cm}, captionskip=-40pt}
    \subfloat[]{\includegraphics[scale=0.67,trim = 0 0 181 0, clip]{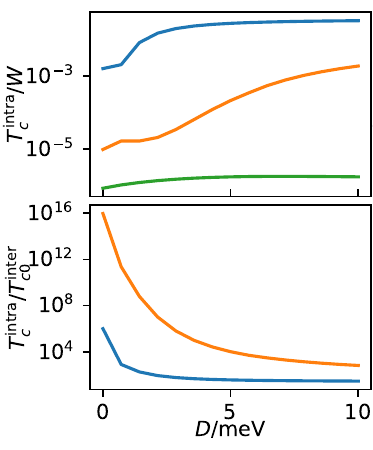}}
    \hfill
    \subfloat[]{\includegraphics[scale=0.67, trim = 180 0 0 0, clip]{figures/two_band_tc_inter_intra.pdf}}
    \vspace{-12pt}
    \caption{Results of the effects of band geometry in the two-band model given in Eq.~\eqref{eq:two_band_Hamiltonian} describing rhombohedral $n$-layer graphene. 
    The top panels of (a) and (b) show the critical temperature of intravalley pairing $T^{\text{intra}}_{c}/W$ and intervalley $T^{\text{inter}}_{c}/W$ respectively for $n = 4, 3, 2$ as a function of the displacement field $D$. The Fermi surface size is kept constant in all the plots. 
    The bottom panels show the same quantities but as a ratio of the critical temperature for intervalley pairing with the pseudospin fully polarized, i.e. the form factor is unity across the band. 
    At small displacement fields we see a big enhancement to $T_c$ that quickly dies out, as the displacement field polarizes the pseudospin. {Since $T_{c0}^{\mathrm{inter}}=0$ for $n=2$, only the results for $n=3$ and $n=4$ are shown.}
    }
    \label{fig:two_band_model_results}
    \vspace{-15pt}
\end{figure}

In Fig.~\ref{fig:two_band_model_results} we show the critical temperature for intravalley and intervalley pairing for $n = 4, \ 3, \ 2$.
In all three cases, we fix the size of the Fermi surface determined by electron filling.
In the top panels of (a) and (b) we plot $T^{\text{intra}}_c/W$, and $T^{\text{inter}}_c/W$ respectively which show an enhancement to the critical temperature as we increase the number of layers. 
This effect is mainly due to the density of states (DOS) increase at the Fermi surface as the dispersion gets flatter. 
Furthermore, there is a general trend for $T_c$ to increase with increasing displacement field, which is again explained by the increase in the DOS at the Fermi surface, since the more gapped the system is the flatter the dispersion. 
In the lower panels (a) and (b) we plot $T^{\text{intra}}_c/T^{\text{inter}}_{c0}$, and $T^{\text{inter}}_c/T^{\text{inter}}_{c0}$ respectively. 
Here $T^{\text{inter}}_{c0}$ is the critical temperature in the case where the pseudospin is fully polarized on either the top or the bottom layer, i.e. $|\mathcal W|=1$. 
The plot shows a large enhancement in the critical temperature due to the band geometry. 
However, this effect quickly diminishes as the displacement field is increased, polarizing the electrons to the bottom layer. 



\section{Discussions}
We show that the electron wave function encoded in the form factor has a significant effect on the critical temperature and pairing symmetry for KL superconductivity. To illustrate the role of the form factor when time-reversal symmetry is broken, we compare $T_c$ with intravalley and intervalley pairing. 
We find that $T_c$ for the intravalley pairing is enhanced compared to the intervalley pairing. {However, strong trigonal warping neglected here may compete with band topology in determining whether intravalley or intervalley pairing is favored, since the $\mathbf{k}$ and $-\mathbf{k}$ states are not at the same energy. \cite{PhysRevB.89.205119}}

For a toy model with LLL form factor, there is an exponential enhancement of $T_c$ depending on the flux enclosed by the Fermi surface. 
Interestingly, $T_c$ develops resonance-like features, depending on the total Berry flux enclosed by the Fermi surface. {We further reveal that it is a resonance between the Berry flux enclosed by the Fermi surface and Cooper pair wave function.}
Instead of being quantized to an integer of unit quantum flux, we analytically show the condition for the optimal $T_c$, which is associated with the maxima of the Bessel functions. 
The resulting pairing symmetry is chiral $p\pm ip$ superconductivity, which hosts a Majorana fermion at vortex cores~\cite{Alicea_2012,Kallin_2016,Read_2000,Sato_2017}. 
The LLL wave function has unique properties in that it has uniform Berry curvature and satisfies the trace condition, where the trace of the quantum metric equals the Berry curvature at all momentum~\cite{Roy_2014,Parameswaran_2013,Wang_2021}. 
The LLL wave function in the flat band limit is ideal for realizing the fractional Chern insulators (FCI)~\cite{Roy_2014,Wu_2012,Ledwith_2020,Mera_2021,PhysRevResearch.6.L032063}, and lots of discussions are devoted to realizing bands with ideal quantum geometry resembling the LLL~\cite{Ledwidth_2022,Wan_2023,Cano_2023}. 
In our work, we reveal that the LLL is ideal for achieving topological superconductivity with high $T_c$ (compared to the Fermi energy). 
$T_c$ is highest when the trace condition is satisfied.


Our results are obtained within the RPA, which involves a geometric summation of bubble diagrams while neglecting other classes of diagrams. 
The RPA is well justified in the limit of a large electron flavor number $N$. 
However, when $N$ is small, as in spin- and valley-polarized metals, it becomes challenging to construct a fully controlled approximation. 
Nevertheless, our RPA analysis clearly demonstrates that band topology and quantum geometry play a significant role in superconductivity. 
Recent unbiased density matrix renormalization group (DMRG) calculations using broadened Landau levels also highlight the importance of band topology in realizing superconducting states~\cite{Wang_2025_b}. 
Developing a more systematically controlled approach applicable at small N remains an open question and an active area of research (see, e.g., Ref.~\cite{Dong_2025}).

The models we have considered so far possess both nonzero Berry curvature and quantum metric. 
In the Supplemental Material~\cite{SM}, we further show, using the RPA method, that superconductivity is also strongly influenced by the single-particle wave function in a model with zero Berry curvature but nonzero quantum geometry~\cite{Hofmann_2022}. 


Our work therefore points to a promising route to achieve high-$T_c$ topological superconductivity through tuning the band geometry. 
Our results suggest that a platform for fractional Chern insulators could also be promising for high $T_c$ topological superconductivity when the band is made dispersive. 
This seems to align with the experimental fact that both FCI and superconductivity have been observed in graphene multilayers.

\acknowledgments
The work is partially supported by the U.S. Department of Energy (DOE) National Nuclear Security Administration (NNSA) under Contract No. 89233218CNA000001 through the Laboratory Directed Research and Development (LDRD) Program and was performed, in part, at the Center for Integrated Nanotechnologies, an Office of Science User Facility operated for the DOE Office of Science, under user Proposals No. 2018BU0010 and No. 2018BU0083.

\emph{Note added}---During the preparation of the
manuscript, we became aware of a related work,
Ref. \cite{Shavit_Alicea_2024}. After completion of the
manuscript, we become aware that the LLL form factor can be derived using a model for rhombohedral graphene~\cite{Bernevig_2025,Jiang_2025}. Valley polarized superconductivity has also been observed in twisted MoTe$_2$ moir\'e superlattice recently, coexisting with FCI in a single device \cite{Xu_Sun2025}, and a Kohn-Luttinger mechanism for superconductivity based on a realistic Chern band was proposed in Ref~\cite{Cheng_2026}.

\bibliography{references}

\setcounter{equation}{0}
\clearpage
\setcounter{figure}{0}
\renewcommand{\theequation}{S\arabic{equation}}

\title{\texorpdfstring{Supplementary materials for ``Enhanced Kohn-Luttinger topological superconductivity in bands with nontrivial geometry"}{}}
\maketitle 

\onecolumngrid
\appendix

\section{Linearized gap equation and the randomized phase approximation}
\begin{figure}[t]
    \centering
    \includegraphics[scale = 0.8]{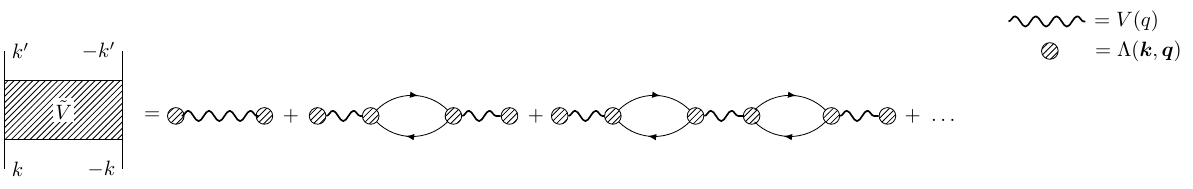}
    \caption{The renormalized interactions under the random phase approximation. We modify every vertex to include the effects of band projection.}
    \label{fig:rpa}
\end{figure}
We follow the approach outlined in Ref.~\cite{Ghazaryan_2021} for the renormalization of the interactions. 
We use a random phase approximation which is controlled by a $1/N$ expansion, $N$ being the number of fermion species, as we sum over the ladder of bubbles as shown in Fig.~\ref{fig:rpa}.
This gives us the renormalized interactions $\mathcal W(\bm k, \bm k') \tilde V(\bm q)$ in the linearized gap equation as written in Eq. (4) in the main text.

To facilitate the numerical evaluation of the eigenvalue problem of the linearized gap equation, we divide the Fermi surface into equal segments and convert the integral equation into a matrix equation, 
\begin{align}
    \Delta_{\bm k} = -\log\left(\frac{W}{T_c} \right) \sum_{\bm k'} \frac{\delta k'}{(2\pi)^2 v_{\bm k'}} \mathcal{W}_{\bm k, \bm k'} \tilde V_{\bm k, \bm k'} \Delta_{\bm k'}
\end{align}
where $\delta k$ is the size of the individual piece on the Fermi surface used in the calculation. 
We further define the normalized order parameter $\bar \Delta_{\bm k} = \sqrt{\frac{\delta k}{v_{\bm k}}} \Delta_{\bm k}$, which gives us the symmetric equation, 
\begin{align}
    \bar \Delta_{\bm k} = - \log\left(\frac{W}{T_c} \right) \sum_{\bm k'}  \frac{1}{(2\pi)^2 } \sqrt{\frac{\delta k \delta k'}{v_{\bm k} v_{\bm k'} } }  \mathcal{W}_{\bm k, \bm k'} \tilde V_{\bm k, \bm k'} \Delta_{\bm k'} = \log\left(\frac{W}{T_c} \right)  \mathcal M_{\bm k, \bm k'} \bar \Delta_{\bm k}
\end{align}
where the matrix $\mathcal M$ is Hermitian. 
It is this matrix that we diagonalize and look for its biggest eigenvalue $\lambda_{\text{max}}$ whose eigenvector has an odd angular momentum for the case of intravalley pairing, or even or odd for the case of intervalley pairing. 

Finally, we comment on our bubble calculation. We use a grid in momentum space whose dimensions is five times as large as the Fermi surface in either direction. 
We divide this interval into a grid of $6000 \times 6000$. 
Ultimately, we need to use a finite temperature, to evaluate the bubble. 
We use a temperature that is of the order of $v_{\bm k} \delta k_{\text{grid}}$, where $\delta k_{\text{grid}}$ is the linear size of the grid. 

\section{Analysis of critical temperature for parabolic dispersion}
\subsection{Lowest Landau level}
In the main text, we showed how we can make analytical progress in the case of large mass. 
Here we give the details of the proof in the main text as well as
show two more cases where analytical progress can be made using the LLL form factor. 

In the main text, we made use of the following identity,
\begin{align}
    e^{\pm iz\sin(\phi)} = J_0(z) + 2 \sum_{n=1}^{\infty} J_{2n}(z) \cos(2n\theta) \pm 2i \sum_{n=1}^{\infty} J_{2n + 1}(z) \sin((2n + 1) \theta), 
\end{align}
so that we have
\begin{align}
    \int \frac{d\phi}{2\pi} \ e^{-il\phi} e^{- iBk_f^2 \sin(\phi)} = (-1)^l J_l(Bk_f^2). 
\end{align}
Focusing on the odd pairing channels, we arrive at the coupling constant $\lambda_l$ given in Eq. (13) of the main text. 

We can also obtain an analytical expression for the critical temperature if we (1) use a parabolic dispersion $\varepsilon(\bm k) = k^2/(2m)$, (2) assume an on-site bare interaction, so that $V(q) = U$, and (3) take $\beta = 0$ in the LLL form factor. 
In this case, we can write 
\begin{align}
    \mathcal V(\phi) = \frac{U e^{-Bk_f^2 (1 - \cos(\phi))}}{1 + \frac{mU}{2\pi} e^{-Bk_f^2 (1 - \cos(\phi))} }
\end{align}
To simplify the notation we define, $\alpha = mU/(2\pi) e^{-Bk_f^2}$, and expand the denominator, 
\begin{align}
    \mathcal V(\phi) = \sum_{n = 1}^{\infty} (-\alpha)^n e^{nBk_f^2 \cos(\phi)}.
\end{align}
We make use of the identity 
\begin{align}
    e^{z \cos(\phi)} = I_0(z) + 2 \sum_{l =1} I_l(z) \cos(l\phi) = \sum_{l=-\infty}^{\infty} I_l(z) e^{il\phi},
\end{align}
where $I_l(z)$ are the modified Bessel functions, and we used the fact that $I_{-l}(z) = I_{l}(z)$ in the last equality. 
We thus arrive at the following expression for the coupling constant at each channel,
\begin{align}
    \lambda_l = - \sum_{n=1}^{\infty} (-\alpha)^n I_{l}(nBk_f^2). 
\end{align}
Even in this case of real $\mathcal W$, we can have attractive channels even using a parabolic dispersion.

Finally, we consider a case where the renormalized interactions decay very rapidly with the momentum transfer $\bm q$ so that we can take 
$\mathcal V(\phi) = \mathcal V_0 \delta(\phi)$. 
In this case we can write, 
\begin{align}
    \lambda_l = - \frac{m}{2\pi} \mathcal V_0 \int \frac{d\phi}{2\pi} = - \frac{m}{2\pi}. 
\end{align}
Thus we have all channels being repulsive. 
Perhaps this is not surprising, since such renormalized interaction has an infinite range in real space, and thus impossible to have an attractive channel. 

\subsection{Two band model for tetralayer graphene}
In the main text, we considered a two-band model that describes $n$-layers of rhombohedral graphene with a displacement field $D$ applied perpendicular to the plane of the sample~\cite{Slizovskiy_2019}.
For this model, we can show that the critical temperature for intervalley and intravalley pairing are equal when $D = 0$, and using the same bubble in both cases. 
For this model we can write down the form factor for $\bm k$ and $\bm k'$ on the Fermi surface as, 
\begin{align}
    \Lambda(\theta, \phi) = \cos^2(\theta/2) + \sin^2(\theta/2) e^{in\phi}, \qquad \bm k, \bm k' \text{on Fermi surface}.
\end{align}
where, $\phi$ is the relative angle between $\bm k$ and $\bm k'$ in the $k_xk_y$-plane, and $ \cos(\theta) = D/\mu $. 
In general, we have 
\begin{align}
    \lambda_l = - \sum_{l'} \mathcal W_{l-l'} \tilde V_{l'}. 
\end{align}
For intravalley pairing, we can write down, 
\begin{align}
    \mathcal W^{\text{intra}} (\theta, \phi) = \left[\Lambda(\theta, \phi)\right]^2 = \cos^4(\theta/2) + 2\sin^2(\theta/2) \cos^2(\theta/2) e^{in\phi} + \sin^4(\theta/2) e^{i2n\phi}
\end{align}
while for intervalley pairing we have 
\begin{align}
    \mathcal W^{\text{inter}} (\phi) = |\Lambda(\theta, \phi) |^2 = \cos^4(\theta/2) +  \sin^4(\theta/2) + \sin^2(\theta/2) \cos^2(\theta/2) (e^{in\phi}  + e^{-in \phi})
\end{align}

For the case when $D=0$, or $\theta = \pi$ we have, 
\begin{align}
    \mathcal W^{\text{intra}} (\phi) =\frac{e^{in\phi}}{4} \left( e^{-in\phi} + 2 + e^{in\phi}\right) = e^{in\phi} \  \mathcal W^{\text{inter}} (\phi)
\end{align}
and thus, 
\begin{align}
    \mathcal W^{\text{intra}}_{l + n} =  \mathcal W^{\text{inter}}_{l}. 
\end{align}
Thus in a situation where $\tilde V_l$ is the same for intervalley and intravalley pairing we have $\lambda^{\text{intra}}_l = \lambda^{\text{inter}}_{l-n}$.
However, in most situations, $\tilde V_l$ differs between the two cases because the polarization bubble in the case of intervalley pairing is double that of the bubble in the case of intervalley pairing. 

\section{zeta lattice results}
\begin{figure}
    \centering
    \includegraphics[]{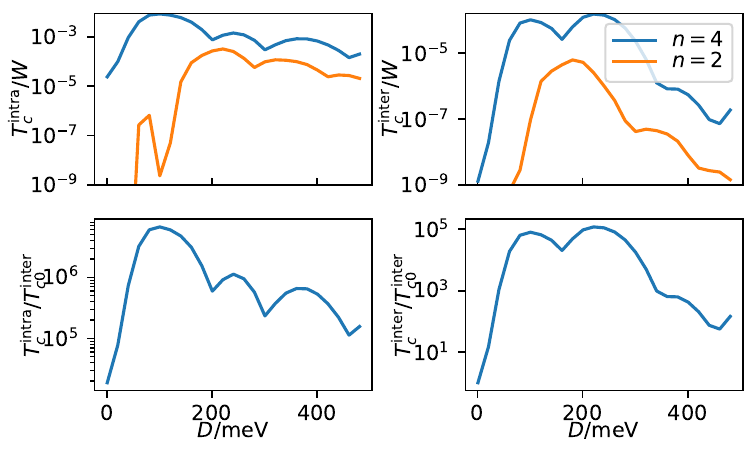}
    \caption{Results of the effects of band projection for the $\zeta$-lattice defined in Ref.~\cite{Hofmann_2022}. $T^{\text{intra}}_c$ is the critical temperature in a valley polarized setting and thus intravalley pairing, while $T^{\text{inter}}_c$ is the critical temperature for intervalley pairing. $T^{\text{inter}}_{c0}$ represents the critical temperature for intervalley coupling with all the vectors are polarized in the same direction on the Fermi surface.}
    \label{fig:zeta_model_results}
\end{figure}

Ref.~\cite{Hofmann_2022} introduced a model with exact flat bands, and a tunable quantum metric, controlled by the parameter $\zeta$. 
The Hamiltonian of such model is given by 
\begin{align}
    &H(\bm k) = -t(\sin\left(\alpha(\bm k)\right) \sigma_x + \cos\left(\alpha(\bm k) \sigma_y \right) \\ 
    &\alpha(\bm k) = \zeta \left(\cos(k_x) + \cos(k_y)\right),
\end{align}
where $\sigma_x$ and $\sigma_y$ are the Pauli matrices. 
This model is interesting in that the Berry curvature vanishes for all points in the Brillouin zone. 
This makes the model interesting in isolating the effects of the quantum metric on $T_c$. 
We have already seen an enhancement is possible for the LLL from factor when $\beta = 0$, corresponding to vanishing Berry curvature. 
Is this effect also true for other models with zero Berry curvature?
We take this model as an additional data point. 
We calculate the form factor using the following expression for the periodic part of Bloch wavefunctions
\begin{align}
    | u (\bm k) \rangle = \frac{1}{\sqrt{2}} \begin{bmatrix} 1 \\ e^{i\alpha(\bm k)} \end{bmatrix}. 
\end{align}
Furthermore, to facilitate the calculation of $T_c$ we further endow each band with a dispersion, 
\begin{align}
    \varepsilon(\bm k) = \gamma \left( \frac{k}{k_0} \right)^n
\end{align}
where $\gamma$ and $k_0$ set the overall scale of energy and momentum respectively. 

In Fig~\ref{fig:zeta_model_results} we show results for the critical temperature for $n = 2, \text{and } 4$. 
In the case of $n=2$, we also see that the effects of band projection allow for superconductivity when it would not be there otherwise. 
We also see resonance peaks in the critical temperature reminiscent to those seen in the LLL case. 
For the $n=4$ dispersion, we comparing the critical temperatures of intravalley and intervalley with band projection, to the expected critical temperature when we ignore the effects of band projection. 
We find enhancement to the critical temperature in both cases. This is unlike the LLL case where the effect of band projection is bad for $T_c$ in the case of intervalley pairing.

\end{document}